\documentclass[prd,reprint,showpacs,showkeys]{revtex4}
\usepackage{amsfonts}
\usepackage{amssymb}
\usepackage{amsmath}
\usepackage{graphicx}
\usepackage[font={footnotesize,it}]{caption}

\begin{document}

\title{Stability of spherically symmetric timelike thin-shells in general
relativity with a variable equation of state}
\author{S. Habib Mazharimousavi}
\email{habib.mazhari@emu.edu.tr}
\author{M. Halilsoy}
\email{mustafa.halilsoy@emu.edu.tr}
\author{S. N. Hamad Amen}
\email{sarbaz256@gmail.com}
\affiliation{Department of Physics, Eastern Mediterranean University, Gazima\~{g}usa,
Turkey. }
\date{\today }

\begin{abstract}
We study spherically symmetric timelike thin-shells in $3+1-$dimensional
bulk spacetime with a variable equation of state for the fluid presented on
the shell. In such a fluid the angular pressure $p$ is a function of both
surface energy density $\sigma $ and the radius $R$ of the thin-shell.
Explicit cases of the thin shells connecting two non-identical cloud of
strings spacetimes and a flat Minkowski spacetime to the Schwarzschild
metric are investigated.
\end{abstract}

\maketitle

\section{Introduction}

Time-like thin-shells in spherically symmetric static spacetimes are the
most interesting cosmological object which can be constructed in general
relativity. Such models of cosmological objects have been used to analyze
some astrophysical phenomena such as gravitational collapse and supernova
explosions. The seminal work of Israel in 1966 \cite{1,2} provided a
concrete formalism for constructing the time-like shells, in general, by
gluing two different manifolds at the location of the thin-shell. As it was
shown in \cite{1}, although the metric tensor of the shell which is induced
by the bulk spacetime presented in both sides of the shell must be
continuous, the extrinsic curvature across the shell is not continuous and
therefore matter has to be introduced on the shell. This formalism has been
employed to study shells in general relativity by many authors for which a
good review paper has been worked out by Kijowski et al in Ref. \cite{3}. In
1990 an exact solution for a static shell which surrounds a black hole was
found by Frauendiener et al \cite{4} and its stability was also studied in
Refs. \cite{5} and \cite{6}. In the formalism introduced by Israel there are
some conditions which are called Israel junction conditions. These
conditions provide a systematic method of finding the energy momentum tensor
presented on the shell. In \cite{7} a computer program was prepared to apply
the junction conditions on the thin-shells in general relativity using
computer algebra. Models of stars and circumstellar shells in general
relativity were studied in \cite{8}. In \cite{9,10} the stability of
spherically symmetric thin-shells was studied while the gravitational
collapse of thin-shells was considered in \cite{11} and \cite{12}.
Thin-shells in Gauss-Bonnet theory of gravity has been studied in \cite{13}
while the rotating thin-shells has been introduced in Ref. \cite{14}.
Stability of charged thin-shells was considered\ in \cite{15} and its
collapse in isotropic coordinates was investigated in \cite{16}. In \cite{17}
charge screening by thin-shells in a $2+1-$dimensional regular black hole
has been studied while the thermodynamics, entropy, and stability of
thin-shells in $2+1-$dimensional flat spacetimes have been given in \cite{18}
and \cite{19}. Recently in \cite{20} the stability of thin-shell interfaces
inside compact stars has been studied by Pereira et al which is very
interesting as they consider a compact star with the core and the crust with
different energy momentum tensors and consequently with different metric
tensors. Screening of the Reissner-Nordstr\"{o}m charge by a thin-shell of
dust matter has also been introduced recently in \cite{21}. Finally one of
the recent works published in this context is about thin-shells joining
local cloud of strings geometries \cite{22}.

In this study first we give a brief review of the thin-shell formalism and
following that we employ a variable equation of state (EoS) for the perfect
fluid presented on the surface of the thin-shell of the form $p=\psi \left(
\sigma ,R\right) $ where $p$ is the surface pressure, $\sigma $ is the
surface energy density and $R$ is the radius of the shell. Let's add that
the function $\psi $ must be differentiable with respect to its arguments $R$
and $\sigma .$ This model of EoS has been introduced first in \cite{23} and
was used recently in \cite{24}. Following the generic stability analysis
with a variable EoS we study two specific cases including a thin-shell
connecting two spacetimes of different cloud of strings and a vacuum flat
Minkowski spacetime connected to a Schwarzschild metric.

\section{Spherically symmetric timelike thin-shells}

In $3+1-$dimensional spherically symmetric bulk spacetime a $2+1-$%
dimensional timelike thin-shell divides the spacetime into two parts which
we shall call as the inside and outside of the shell. The line elements of
the spacetimes in different sides must naturally be different otherwise the
thin-shell becomes a trivial invisible object. Let's label the spacetime
inside the shell as $1$ and outside the shell as $2$. Hence the line element
of each side may be written generically as%
\begin{equation}
ds_{a}^{2}=-f_{a}\left( r_{a}\right) dt_{a}^{2}+\frac{dr_{a}^{2}}{%
f_{a}\left( r_{a}\right) }+r_{a}^{2}\left( d\theta _{a}^{2}+\sin ^{2}\theta
_{a}d\varphi _{a}^{2}\right)
\end{equation}%
in which $a=1,2$ for inside and outside, respectively. We add that in
general the coordinates i.e., $t_{a},r_{a},\theta _{a}$ and $\varphi _{a}$
need not be the same. In general a thin-shell is a constraint condition on
the coordinates of the bulk spacetime but in our study the thin-shell is
defined by $F:=r_{a}-R\left( \tau \right) =0$ in which $\tau $ is the proper
time measured by an observer on the shell such that on both sides we define%
\begin{equation}
-f_{a}\left( R\right) \left( \frac{dt_{a}}{d\tau }\right) ^{2}+\frac{1}{%
f_{a}\left( R\right) }\left( \frac{dR}{d\tau }\right) ^{2}=-1.
\end{equation}%
As one of the Israel junction condition, $ds_{\left( ts\right) }^{2}$ from
one side to the other of the thin-shell must be continuous. Hence, the
coordinates on the shell i.e., $\tau ,\theta _{a}$ and $\varphi _{a}$ have
to be identical on both sides so that we shall remove the sub index $a.$
This results in a unique induced metric on the shell which is applicable to
both sides expressed by%
\begin{equation}
ds_{\left( ts\right) }^{2}=-d\tau ^{2}+R^{2}\left( \tau \right) \left(
d\theta ^{2}+\sin ^{2}\theta d\varphi ^{2}\right) .
\end{equation}%
Before we proceed further let us note that although the proper time in
different sides of the shell is common the coordinate times $t_{a}$ are
different and they are found from (3) as 
\begin{equation}
\dot{t}_{a}^{2}=\frac{f_{a}\left( R\right) +\dot{R}^{2}}{f_{a}^{2}\left(
R\right) }
\end{equation}%
in which a dot stands for the derivative with respect to the proper time $%
\tau $. Let's also add that $t_{1}$ and $t_{2}$ refer to the coordinate
times of the inner and outer regions.

For future use we set our coordinate systems of the bulk i.e., $%
ds_{a}^{2}=g_{\mu \nu }^{\left( a\right) }dx^{\left( a\right) \mu
}dx^{\left( a\right) \nu }$ and the shell i.e., $ds_{\left( ts\right)
}^{2}=h_{ij}d\xi ^{i}d\xi ^{j}$ as follow: for the bulk spacetime $x^{\left(
a\right) \mu }=\left\{ t_{a},r_{a},\theta _{a},\varphi _{a}\right\} $ and $%
\xi ^{i}=\left\{ \tau ,\theta ,\varphi \right\} $ for the thin-shell. The
second fundamental form (or extrinsic curvature) tensor of the shell in each
side can be found as%
\begin{equation}
K_{ij}^{\left( a\right) }=-n_{\gamma }^{\left( a\right) }\left( \frac{%
\partial ^{2}x^{\left( a\right) \gamma }}{\partial \xi ^{i}\partial \xi ^{j}}%
+\Gamma _{\alpha \beta }^{\left( a\right) \gamma }\frac{\partial x^{\left(
a\right) \alpha }}{\partial \xi ^{i}}\frac{\partial x^{\left( a\right) \beta
}}{\partial \xi ^{j}}\right)
\end{equation}%
in which $n_{\gamma }^{\left( a\right) }$ is the four-normal spacelike
vector on each side of the thin-shell pointing outward given by 
\begin{equation}
n_{\gamma }^{\left( a\right) }=\left( -\dot{R}\left( \tau \right) ,\dot{t}%
_{a},0,0\right) .
\end{equation}%
The nonzero components of the extrinsic curvature, therefore, are found to be%
\begin{equation}
K_{\tau \tau }^{\left( a\right) }=-\frac{2\ddot{R}\left( \tau \right)
+f_{a}^{\prime }\left( R\right) }{2\sqrt{f_{a}\left( R\right) +\dot{R}^{2}}},
\end{equation}%
\begin{equation}
K_{\theta \theta }^{\left( a\right) }=R\left( \tau \right) \sqrt{f_{a}\left(
R\right) +\dot{R}^{2}}
\end{equation}%
and%
\begin{equation}
K_{\varphi \varphi }^{\left( a\right) }=R\left( \tau \right) \sqrt{%
f_{a}\left( R\right) +\dot{R}^{2}}\sin ^{2}\theta .
\end{equation}%
One observes that unlike the first fundamental form, the second fundamental
form is not continuous in general. However, if $f_{a}\left( R\right) $ and $%
f_{a}^{\prime }\left( R\right) $ are the same on both sides of the shell
then $K_{ij}$ as well as $h_{ij}$ are both continuous. In case that $K_{ij}$
is not continuous it satisfies the other Israel junction condition%
\begin{equation}
\lbrack K_{i}^{j}]-\delta _{i}^{j}\left[ K\right] =-8\pi GS_{i}^{j}
\end{equation}%
in which $[K_{i}^{j}]=K_{i}^{\left( 2\right) j}-$ $K_{i}^{\left( 1\right)
j}, $ $\left[ K\right] =trac[K_{i}^{j}]=[K_{i}^{i}]$ and 
\begin{equation}
S_{i}^{j}=diag\left( -\sigma ,p,p\right)
\end{equation}%
is the energy-momentum tensor of the thin-shell. Herein, $\sigma $ is the
energy density and $p$ the angular pressure on the shell. We note that, as
we have considered the bulk to be spherically symmetric, the pressures in $%
\theta $ and $\varphi $ directions are identical and the energy-momentum
tensor is of a perfect fluid type. The explicit form of the $\sigma $ and $p$
are given by 
\begin{equation}
\sigma =-\frac{1}{4\pi G}\left( \frac{\sqrt{f_{2}\left( R\right) +\dot{R}^{2}%
}-\sqrt{f_{1}\left( R\right) +\dot{R}^{2}}}{R\left( \tau \right) }\right) ,
\end{equation}%
and%
\begin{equation}
p=\frac{1}{8\pi G}\left( \frac{2\ddot{R}\left( \tau \right) +f_{2}^{\prime
}\left( R\right) }{2\sqrt{f_{2}\left( R\right) +\dot{R}^{2}}}-\frac{2\ddot{R}%
\left( \tau \right) +f_{1}^{\prime }\left( R\right) }{2\sqrt{f_{1}\left(
R\right) +\dot{R}^{2}}}+\frac{\sqrt{f_{2}\left( R\right) +\dot{R}^{2}}-\sqrt{%
f_{1}\left( R\right) +\dot{R}^{2}}}{R\left( \tau \right) }\right) .
\end{equation}%
In the case of a static thin-shell one has to set $R\left( \tau \right)
=R_{0}$ which is a constant and consequently 
\begin{equation}
\sigma _{0}=-\frac{1}{4\pi G}\left( \frac{\sqrt{f_{2}\left( R_{0}\right) }-%
\sqrt{f_{1}\left( R_{0}\right) }}{R_{0}}\right) ,
\end{equation}%
and%
\begin{equation}
p_{0}=\frac{1}{8\pi G}\left( \frac{f_{2}^{\prime }\left( R_{0}\right) }{2%
\sqrt{f_{2}\left( R_{0}\right) }}-\frac{f_{1}^{\prime }\left( R_{0}\right) }{%
2\sqrt{f_{1}\left( R_{0}\right) }}+\frac{\sqrt{f_{2}\left( R_{0}\right) }-%
\sqrt{f_{1}\left( R_{0}\right) }}{R_{0}}\right) .
\end{equation}

\section{Stability}

Let's assume that our constructed thin-shell is at equilibrium at $R=R_{0}$
which means $\dot{R}=\ddot{R}=0$ and therefore the energy density and the
pressures are given by Eqs. (14) and (15). Any radial perturbation causes
the radius of the shell to be changed in a dynamical sense. In other words
after the radial perturbation $R$ becomes a function of proper time $\tau $
and consequently the energy density and the angular pressures are found to
be Eqs. (12) and (13) which also satisfy%
\begin{equation}
\frac{d\sigma }{dR}+\frac{2}{R}\left( p+\sigma \right) =0.
\end{equation}%
This equation is the dynamic relation that connects $p$ and $\sigma $ after
the perturbation. Furthermore, any kind of fluid presented on the shell has
to satisfy an equation of state which is nothing but a relation between $p$
and $\sigma .$ This relation is traditionally expressed as 
\begin{equation}
p=p\left( \sigma \right)
\end{equation}%
but in our study we use a more general EoS given by \cite{23,24}%
\begin{equation}
p=\psi \left( R,\sigma \right) .
\end{equation}%
A substitution in (16) yields%
\begin{equation}
\frac{d\sigma }{dR}+\frac{2}{R}\left( \psi \left( R,\sigma \right) +\sigma
\right) =0
\end{equation}%
which is a principal equation that connects $\sigma $ to $R$ after the
perturbation. In addition to this, from the explicit form of $\sigma $ in
Eq. (12) we find%
\begin{equation}
\dot{R}^{2}+V\left( R,\sigma \left( R\right) \right) =0
\end{equation}%
in which%
\begin{equation}
V\left( R,\sigma \left( R\right) \right) =\frac{f_{1}\left( R\right)
+f_{2}\left( R\right) }{2}-\frac{\left( f_{1}\left( R\right) -f_{2}\left(
R\right) \right) ^{2}}{\left( 8\pi GR\sigma \left( R\right) \right) ^{2}}%
-\left( 2\pi GR\sigma \left( R\right) \right) ^{2}.
\end{equation}%
This equation is a one dimensional equation of motion for the radius of the
thin-shell after the perturbation. Together with Eq. (19) gives a clear
picture of the motion of the thin-shell after the perturbation. More
precisely, the solution of Eq. (19) is used in (20) and the general motion
of the radius of the thin-shell, in principle is found by solving Eq. (20).
The nature of the motion after the perturbation depends on the form of the
function $\psi \left( R,\sigma \right) $ given by EoS and the metric
functions $f_{1}\left( R\right) $ and $f_{2}\left( R\right) .$ We comment
that the general one-dimensional equation of motion (20) is highly
non-linear. In general we do not expect an exact, closed form solution for
the radius of the thin-shell after the perturbation. A linearized version of
this equation helps us to know the general behaviour of the motion of the
thin-shell after the perturbation without going through the complete
solution. As we have stated the thin-shell is in equilibrium at $R=R_{0}$ so
we expand $V\left( R,\sigma \left( R\right) \right) $ about $R=R_{0}$ and
keep it at the first nonzero term. This is called a linearized radial
perturbation. The expansion of $V\left( R,\sigma \left( R\right) \right) $
about $R=R_{0}$ reads as 
\begin{equation}
V\left( R,\sigma \right) =V\left( R_{0},\sigma _{0}\right) +\left. \frac{dV}{%
dR}\right\vert _{R=R_{0}}\left( R-R_{0}\right) +\frac{1}{2}\left. \frac{%
d^{2}V}{dR^{2}}\right\vert _{R=R_{0}}\left( R-R_{0}\right) ^{2}+\mathcal{O}%
\left( \left( R-R_{0}\right) ^{3}\right)
\end{equation}%
where $V\left( R_{0},\sigma _{0}\right) $ and $\left. \frac{dV}{dR}%
\right\vert _{R=R_{0}}$ both are zero identically i.e., the first because $%
\dot{R}_{0}^{2}=0$ and the second because $R=R_{0}$ is the equilibrium
radius in the sense that the force is zero there. Introducing $x=R-R_{0},$
up to the second order we get%
\begin{equation}
\dot{x}^{2}+\omega ^{2}x^{2}\simeq 0
\end{equation}%
in which%
\begin{equation}
\omega ^{2}=\frac{1}{2}\left. \frac{d^{2}V}{dR^{2}}\right\vert _{R=R_{0}}.
\end{equation}%
Derivative with respect to $\tau $ implies%
\begin{equation}
\ddot{x}+\omega ^{2}x\simeq 0
\end{equation}%
which clearly for $\omega ^{2}>0$ represents an oscillation about $x=0.$
This, however, means the radius of the thin-shell oscillates about the
equilibrium radius $R=R_{0}.$ This is what we mean by a stable condition. In
other words if 
\begin{equation}
\frac{1}{2}\left. \frac{d^{2}V}{dR^{2}}\right\vert _{R=R_{0}}>0
\end{equation}%
the thin-shell oscillates and remains stable. Unlike $\omega ^{2}>0$ if $%
\omega ^{2}<0$ then the motion of the radius of the thin-shell becomes an
exponential form which yields an unstable thin-shell. To proceed further we
need to calculate $\left. \frac{d^{2}V}{dR^{2}}\right\vert _{R=R_{0}}$ and
as $V=V\left( R,\sigma \right) $ we shall need $\sigma ^{\prime }=\frac{%
d\sigma }{dR}$ and $\sigma ^{\prime \prime }=\frac{d^{2}\sigma }{dR^{2}}.$
Upon differentiation with respect to $R,$ (19) yields%
\begin{equation}
\sigma ^{\prime \prime }=\frac{2}{R^{2}}\left( \psi +\sigma \right) \left( 2%
\frac{\partial \psi }{\partial \sigma }+3\right) -\frac{2}{R}\frac{\partial
\psi }{\partial R}.
\end{equation}%
Next, we find $V^{\prime \prime }\left( R\right) $ at $R=R_{0}$ by applying $%
\sigma ^{\prime }$ and $\sigma ^{\prime \prime }$ whenever we need them
which becomes%
\begin{multline}
V_{0}^{\prime \prime }=-\frac{16\pi GF_{0}H_{0}}{F_{0}-H_{0}}\psi _{,R}+%
\frac{2\left[ H_{0}\left( 2F_{0}^{2}-f_{10}^{\prime }R_{0}\right)
-F_{0}\left( 2H_{0}^{2}-f_{20}^{\prime }R_{0}\right) \right] }{\left(
F-H\right) R_{0}^{2}}\psi _{,\sigma }+ \\
\frac{\left[ 4F_{0}^{4}-2R_{0}\left( f_{10}^{\prime }+R_{0}f_{10}^{\prime
\prime }\right) F_{0}^{2}+R_{0}^{2}f_{10}^{\prime 2}\right] H_{0}^{3}-\left[
4H_{0}^{4}-2R_{0}\left( f_{20}^{\prime }+R_{0}f_{20}^{\prime \prime }\right)
H_{0}^{2}+R_{0}^{2}f_{20}^{\prime 2}\right] F_{0}^{3}}{2\left(
F_{0}-H_{0}\right) F_{0}^{2}H_{0}^{2}R_{0}^{2}}
\end{multline}%
in which $F_{0}=\sqrt{f_{10}},$ $H_{0}=\sqrt{f_{20}},\psi _{,R}=\left. \frac{%
\partial \psi }{\partial R}\right\vert _{R=R_{0}}$ and $\psi _{,\sigma
}=\left. \frac{\partial \psi }{\partial \sigma }\right\vert _{R=R_{0}}.$

\section{Thin-shell connecting two spacetimes of cloud of strings}

In this section we consider $f_{1}=\kappa _{1}$ and $f_{2}=\kappa _{2}$ in
which $\kappa _{1}$ and $\kappa _{2}$ are two positive constants satisfying $%
\kappa _{1}>\kappa _{2}$. Such spacetimes where $\kappa _{1},\kappa _{2}\neq
1$ represent the so called cloud of strings spacetime \cite{22,25,26,27,28}.
The energy momentum tensor at the equilibrium is given by%
\begin{equation}
S_{i}^{j}=-\frac{\sqrt{\kappa _{1}}-\sqrt{\kappa _{2}}}{4\pi GR_{0}}%
diag\left( 1,-\frac{1}{2},-\frac{1}{2}\right) 
\end{equation}%
which means%
\begin{equation}
\sigma _{0}=\frac{\sqrt{\kappa _{1}}-\sqrt{\kappa _{2}}}{4\pi GR_{0}}\text{ }
\end{equation}%
and%
\begin{equation}
p_{0}=-\frac{1}{2}\sigma _{0}.
\end{equation}%
This energy momentum tensor satisfies the weak energy condition which states
that $\sigma _{0}\geq 0,$ and $\sigma _{0}+p_{0}\geq 0$ and therefore it is
physical. To proceed with the stability analysis we must choose an EoS. For
the first choice we set 
\begin{equation}
\frac{\partial \psi }{\partial R}=\omega _{1}
\end{equation}%
and%
\begin{equation}
\frac{\partial \psi }{\partial \sigma }=\omega _{2}
\end{equation}%
in which both $\omega _{1}$ and $\omega _{2}$ are constants. The general
form of $V^{\prime \prime }\left( R_{0}\right) $ given in Eq. (28) yields%
\begin{equation}
V^{\prime \prime }\left( R_{0}\right) =-\frac{2\sqrt{\kappa _{1}\kappa _{2}}%
\left[ 8G\pi \omega _{1}R_{0}^{2}-\left( 2\omega _{2}+1\right) \left( \sqrt{%
\kappa _{1}}-\sqrt{\kappa _{2}}\right) \right] }{R_{0}^{2}\left( \sqrt{%
\kappa _{1}}-\sqrt{\kappa _{2}}\right) }.
\end{equation}%
In order to have $V^{\prime \prime }\left( R_{0}\right) >0$ one must impose 
\begin{equation}
8G\pi \omega _{1}R_{0}^{2}-\left( 2\omega _{2}+1\right) \left( \sqrt{\kappa
_{1}}-\sqrt{\kappa _{2}}\right) <0
\end{equation}%
which in turn implies that 
\begin{equation}
\omega _{1}<\frac{\left( \sqrt{\kappa _{1}}-\sqrt{\kappa _{2}}\right) }{%
8G\pi R_{0}^{2}}\left( 2\omega _{2}+1\right) .
\end{equation}%
The case where $\omega _{1}=0$ implies a linear perfect fluid which is
stable for $\omega _{2}>-\frac{1}{2}$ and is unstable for $\omega _{2}<-%
\frac{1}{2}.$ For the case where $\omega _{1}\neq 0$ the situation becomes
more complicated. In Fig. 1 we plot 
\begin{equation}
\omega _{1}=\frac{\left( \sqrt{\kappa _{1}}-\sqrt{\kappa _{2}}\right) }{%
8G\pi R_{0}^{2}}\left( 2\omega _{2}+1\right) 
\end{equation}%
for $\frac{\left( \sqrt{\kappa _{1}}-\sqrt{\kappa _{2}}\right) }{8G\pi
R_{0}^{2}}=0.1,0.2,0.3$ and $0.4.$ As it is imposed from the condition (35)
the values of $\omega _{1}$ and $\omega _{2}$ under the lines for each
specific choice of $\frac{\left( \sqrt{\kappa _{1}}-\sqrt{\kappa _{2}}%
\right) }{8G\pi R_{0}^{2}}$ yields a region of stability while the opposite
side (above the lines) stands for the values of $\omega _{1}$ and $\omega
_{2}$ that result in an unstable thin-shell. In Ref. \cite{22} thin-shells
joining local cloud of strings geometries has been studied by Eiroa et al.
Although their line elements for the cloud of strings geometries was written
in static axially symmetric form and their equation of state was linear
perfect fluid, their results is in agreement with what we found here in Eq.
(36) provided $\omega _{1}=0.$ Let's add that $\omega _{1}<0$ increases the
region of stability in comparison with the linear perfect fluid with $\omega
_{1}=0.$ This can be seen from Eq. (36) and in Fig. 1.

\begin{figure}[h]
\includegraphics[width=120mm,scale=0.7]{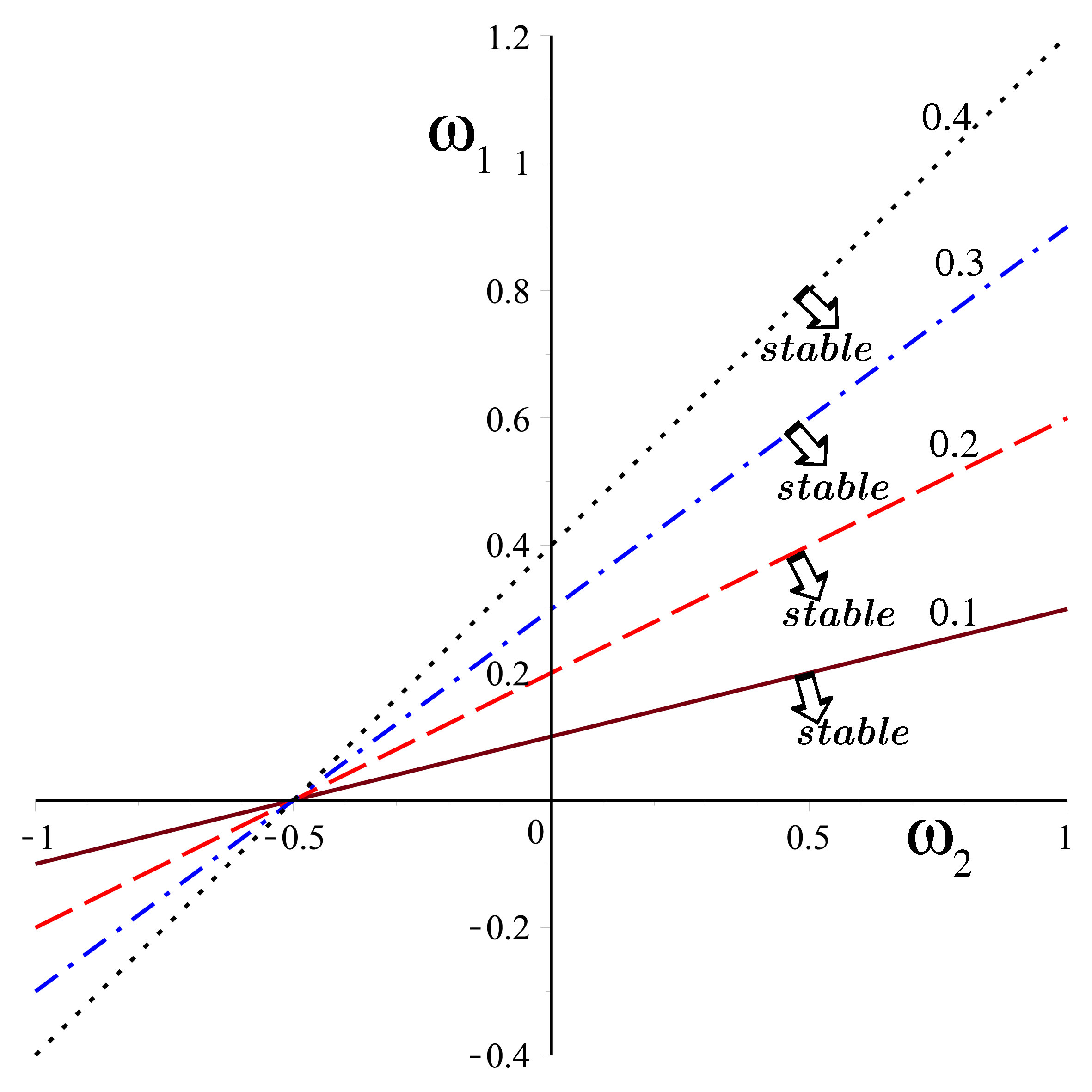}
\caption{A plot of $\protect\omega _{1}$ with respect to $\protect\omega %
_{2} $ for various values of $\frac{\left( \protect\sqrt{\protect\kappa _{1}}%
-\protect\sqrt{\protect\kappa _{2}}\right) }{8G\protect\pi R_{0}^{2}}%
=0.1,0.2,0.3$ and $0.4.$ The arrows show the region of stability while the
opposite side is the unstable zone for each case. }
\label{fig: 1}
\end{figure}

\section{Thin-shell connecting vacuum to Schwarzschild}

In this part we consider the inner spacetime to be flat with $f_{1}=1$ and
the outer spacetime the Schwarzschild with $f_{2}\left( r_{2}\right) =1-%
\frac{2m}{r_{2}}.$ The closed forms of $\sigma _{0}$ and $p_{0}$ are found
to be%
\begin{equation}
\sigma _{0}=\frac{1-\sqrt{1-\frac{2m}{R_{0}}}}{4\pi GR_{0}}
\end{equation}%
and%
\begin{equation}
p_{0}=-\frac{m-R_{0}+R_{0}\sqrt{1-\frac{2m}{R_{0}}}}{8\pi GR_{0}^{2}\sqrt{1-%
\frac{2m}{R_{0}}}}.
\end{equation}%
These clearly satisfy the weak energy conditions i.e., $\sigma _{0}\geq 0$
and $\sigma _{0}+p_{0}\geq 0$ provided $R_{0}>2m.$ Furthermore, one finds%
\begin{equation}
V_{0}^{\prime \prime }=\frac{16G\pi \Omega }{\Omega -1}\psi _{,R}+\frac{%
2\left( 2\psi _{,\sigma }+1\right) \Omega }{R_{0}^{2}}
\end{equation}%
in which $\Omega =\sqrt{1-\frac{2m}{R_{0}}}>0$, $\psi _{,R}=\left. \frac{%
\partial \psi }{\partial R}\right\vert _{R=R_{0}}$ and $\psi _{,\sigma
}=\left. \frac{\partial \psi }{\partial \sigma }\right\vert _{R=R_{0}}.$ In
this case also we set $\frac{\partial \psi }{\partial R}=\omega _{1}$ and $%
\frac{\partial \psi }{\partial \sigma }=\omega _{2}$ which implies 
\begin{equation}
V_{0}^{\prime \prime }=-\frac{16G\pi \Omega }{1-\Omega }\omega _{1}+\frac{%
2\left( 2\omega _{2}+1\right) \Omega }{R_{0}^{2}}.
\end{equation}%
For the case $\omega _{1}=0$ which corresponds to a linear perfect fluid one
finds $V_{0}^{\prime \prime }\geq 0$ with $2\omega _{2}+1\geq 0$ or
equivalently $\omega _{2}\geq -\frac{1}{2}.$ For the case $\omega _{1}\neq 0$
we have to work out the regions in the plane of $\omega _{1}$ and $\omega
_{2}$ such that $V_{0}^{\prime \prime }\geq 0.$ To find the region where $%
V_{0}^{\prime \prime }\geq 0$ we find $\omega _{2}$ in terms of $\omega _{1}$
such that $V_{0}^{\prime \prime }=0$. This amounts to 
\begin{equation}
\omega _{2}=-\frac{1}{2}+\chi \omega _{1}
\end{equation}%
in which%
\begin{equation}
\chi =\frac{4\pi GR_{0}^{2}}{1-\sqrt{1-\frac{2m}{R_{0}}}}.
\end{equation}%
Depending on the value of $m$ and $R_{0}>2m,$ one finds%
\begin{equation}
4\pi GR_{0}^{2}<\chi <\infty .
\end{equation}%
In Fig. 2 we plot $\omega _{2}$ versus $\omega _{1}$ for various values for $%
\chi =0.1,0.2,0.3$ and $0.4.$ Also the stability regions for each case is
shown by an arrow indicator. We would like to add that such a thin-shell has
been indirectly considered by Eiroa and Simeone in \cite{15} with a linear
perfect fluid. In both Fig. 1 and 2 in \cite{15}, the upper left, represents
the thin-shell connecting the vacuum to Schwarzschild with their parameter $%
\eta $. Presence of $\omega _{1}$ changes the stability regime, especially
if $\omega _{1}<0,$ (41) implies that the term due to $\omega _{1}$ is
positive which increases the stability region.

\begin{figure}[h]
\includegraphics[width=120mm,scale=0.7]{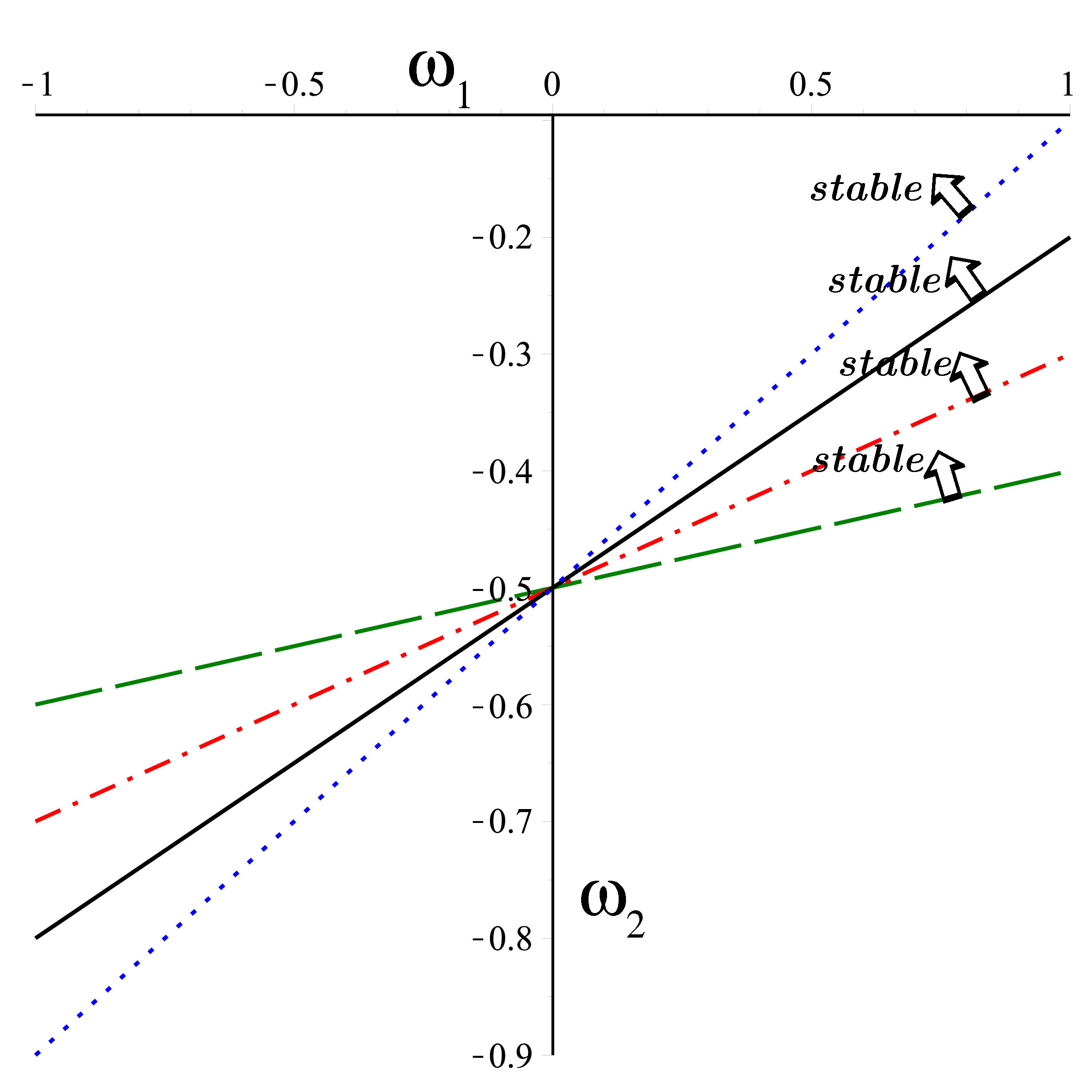}
\caption{A plot of $\protect\omega _{2}$ with respect to $\protect\omega %
_{1} $ for various values of $\protect\chi =0.1,0.2,0.3$ and $0.4.$ The
arrows show the region of stability while the opposite side is the unstable
zone for each case. }
\end{figure}

Before completing this section we would like to find the explicit form of
the energy density $\sigma $ after the perturbation. In both examples we
have worked out in this chapter we assumed $\frac{\partial \psi }{\partial R}%
=\omega _{1}$ and $\frac{\partial \psi }{\partial \sigma }=\omega _{2}$ in
which $\omega _{1}$ and $\omega _{2}$ are two constants. Integration with
respect to $R$ and $\sigma $ results in%
\begin{equation}
\psi =\omega _{1}R+\omega _{2}\sigma +C_{0}
\end{equation}%
in which $C_{0}$ is an integration constant. As $p=\psi $ should give the
equilibrium pressure at $R=R_{0}$ we can find the value of $C_{0}$ as 
\begin{equation}
C_{0}=p_{0}-\omega _{1}R_{0}-\omega _{2}\sigma _{0}
\end{equation}%
and therefore the dynamic pressure becomes%
\begin{equation}
p=\omega _{1}\left( R-R_{0}\right) +\omega _{2}\left( \sigma -\sigma
_{0}\right) +p_{0}.
\end{equation}%
This EoS together with Eq. (19) gives the differential equation 
\begin{equation}
\frac{d\sigma }{dR}+\frac{2}{R}\left( \omega _{1}\left( R-R_{0}\right)
+\omega _{2}\left( \sigma -\sigma _{0}\right) +p_{0}+\sigma \right) =0
\end{equation}%
which must be satisfied by $\sigma .$ The solution of this equation is given
by%
\begin{equation}
\sigma \left( R\right) =\frac{\omega _{2}\sigma _{0}-p_{0}+\omega _{1}R_{0}}{%
1+\omega _{2}}-\frac{2\omega _{1}R}{3+2\omega _{2}}+\frac{C_{1}}{R^{2\left(
\omega _{2}+1\right) }}
\end{equation}%
in which $C_{1}$ is an integration constant. Imposing $\sigma \left(
R_{0}\right) =\sigma _{0}$ yields%
\begin{equation}
C_{1}=R_{0}^{2\left( \omega _{2}+1\right) }\left( \frac{\sigma _{0}+p_{0}}{%
1+\omega _{2}}-\frac{\omega _{1}R_{0}}{3+5\omega _{2}+2\omega _{2}^{2}}%
\right) .
\end{equation}%
Finally the closed form of the energy density is found to be%
\begin{equation}
\sigma \left( R\right) =\frac{\omega _{2}\sigma _{0}-p_{0}+\omega _{1}R_{0}}{%
1+\omega _{2}}-\frac{2\omega _{1}R}{3+2\omega _{2}}+\left( \frac{R_{0}}{R}%
\right) ^{2\left( \omega _{2}+1\right) }\left( \frac{\sigma _{0}+p_{0}}{%
1+\omega _{2}}-\frac{\omega _{1}R_{0}}{3+5\omega _{2}+2\omega _{2}^{2}}%
\right) .
\end{equation}%
We note that at $R=R_{0},$ $\sigma \left( R\right) $ reduces to $\sigma _{0}$
and it is a function of $R$ as well as $\omega _{1}$ and $\omega _{2}.$ The
case $\omega _{1}=0$ admits%
\begin{equation}
\sigma \left( R\right) =\frac{\omega _{2}\sigma _{0}-p_{0}}{1+\omega _{2}}%
+\left( \frac{R_{0}}{R}\right) ^{2\left( \omega _{2}+1\right) }\left( \frac{%
\sigma _{0}+p_{0}}{1+\omega _{2}}\right)
\end{equation}%
while when $\omega _{2}=0$ we find%
\begin{equation}
\sigma \left( R\right) =-p_{0}+\omega _{1}R_{0}-\frac{2\omega _{1}R}{3}%
+\left( \frac{R_{0}}{R}\right) ^{2}\left( \sigma _{0}+p_{0}-\frac{\omega
_{1}R_{0}}{3}\right) .
\end{equation}%
In the case both $\omega _{1}$ and $\omega _{2}$ are set to zero the energy
density becomes 
\begin{equation}
\sigma \left( R\right) =-p_{0}+\left( \frac{R_{0}}{R}\right) ^{2}\left(
\sigma _{0}+p_{0}\right)
\end{equation}%
while 
\begin{equation}
p=p_{0}
\end{equation}%
even after the perturbation. According to the Fig. 2, this is one of the
case where the thin-shell is stable with 
\begin{equation}
V_{0}^{\prime \prime }=\frac{2\Omega }{R_{0}^{2}}
\end{equation}%
which is clearly positive.

\section{CONCLUSION}

We studied the stability of the timelike thin-shells in spherically
symmetric spacetimes with a variable EoS on the shell of the form $p=\psi
\left( \sigma ,R\right) $, \cite{23,24}. We recall that in previous studies
this was taken simply as $p=p\left( \sigma \right) .$ We analyzed the
stability of the thin-shells against a radial perturbation and by a
linearized approximation we found a general condition to be satisfied in
order to have a stable spherically symmetric thin-shell. We applied our
results to two explicit cases with the variable EoS on the shell and
numerically, as well as analytically we obtained the stability regions. Our
first thin-shell joins two spacetimes of local cloud of strings,
characterized by two distinct parameters $\kappa _{1}>0$ and $\kappa _{2}>0$
such that $\kappa _{1}>$ $\kappa _{2}$. The second thin-shell connects a
vacuum flat spacetime to the Schwarzschild metric. In both examples the weak
energy conditions are satisfied for which the stability regions are
identified and plotted.


\begin{thebibliography}{99}
\bibitem{1} W. Israel. Nuovo Cimento B, 44:1, (1966).

\bibitem{2} W. Israel. Nuovo Cimento B, 48:463, (1966).

\bibitem{3} D. Kijowski, J. Magli. Gen. Relativ. Gravitation, 38:1697,
(2006).

\bibitem{4} C. Frauendiener, J. Hoenselaers and W. Konrad. Classical Quantum
Gravity, 7:583, (1990).

\bibitem{5} J. Brady, P. R. Louko and E. Poisson. Phys. Rev. D, 44:1891,
(1991).

\bibitem{6} B. G. Schmidt. Phys. Rev. D, 59:024005, (1998).

\bibitem{7} P. Musgrave and K. Lake. Class.Quant.Grav., 13:1885, (1996).

\bibitem{8} K. G. Zloshchastiev. Int.J.Mod.Phys. D, 8:549, (1999).

\bibitem{9} M. Ishak and K. Lake. Phys. Rev. D, 65:044011, (2002).

\bibitem{10} F. S. N. Lobo and P. Crawford. Class. Quant. Grav., 22:4869,
(2005).

\bibitem{11} J. Crisostomo and R. Olea. Phys. Rev. D, 69:104023, (2004).

\bibitem{12} S. del. Crisostomo, J. Campo and J. Saavedra. Phys. Rev. D,
70:064034, (2004).

\bibitem{13} E. Gravanis and S. Willison. Phys. Rev. D, 75:084025, (2007).

\bibitem{14} R. J. Gleiser and M. A. Ramirez. Class. Quant. Grav.,
26:045006, (2009).

\bibitem{15} E. F. Eiroa and C. Simeone. Phys. Rev. D, 83:104009, (2011).

\bibitem{16} H. Beauchesne and A. Edery. Phys. Rev. D, 85:044056, (2012).

\bibitem{17} S. H. Mazharimousavi and M. Halilsoy. Eur. Phys. J. C.,
73:2527, (2013).

\bibitem{18} J. P. S. Lemos and G. M. Quinta. Phys. Rev. D, 88:067501,
(2013).

\bibitem{19} J. P. S. Lemos and G. M. Quinta. Phys. Rev. D, 89:084051,
(2014).

\bibitem{20} J. G. Pereira, J. P. Coelho and J. A. Rueda. Phys. Rev. D,
90:123011, (2014).

\bibitem{21} S. H. Mazharimousavi and M. Halilsoy. Eur. Phys. J. C., 75:334,
(2015).

\bibitem{22} E. Eiroa, E. F. Rubin de Celis and C. Simeone. Eur. Phys. J.
C., 76:546, (2016).

\bibitem{23} N. M. Garcia, F. S. N. Lobo and M. Visser, Phys. Rev. D
86:044026, (2012).

\bibitem{24} V. Varela Phys. Rev. D 92:044002, (2015).

\bibitem{25}  P. S. Letelier, Phys. Rev. D 20:1294 (1979). 

\bibitem{26}  S. G. Ghosh and S. D. Maharaj, Phys. Rev. D 89:084027 (2014).

\bibitem{27}  S. G. Ghosh, U. Papnoi and S. D. Maharaj, Phys. Rev. D
90:044068 (2014).

\bibitem{28} S. H. Mazharimousavi and M. Halilsoy, Eur. Phys. J. C 76:95
(2016).
\end{thebibliography}
\end{document}